\documentclass
[twocolumn,showpacs,showkeys,preprintnumbers,prl,superscriptaddress]{revtex4}%
\usepackage{graphics}
\usepackage{amsmath}
\usepackage{amsfonts}
\usepackage{amssymb}
\usepackage{graphicx}%
\begin{document}
\title{Stochastic Annealing}
\date{\today}

%authors from authors.tex

\author{Robin C. Ball}
\email{r.c.ball@warwick.ac.uk}
\homepage{http://www.phys.warwick.ac.uk/theory/}
\affiliation{Department of Physics, University of Warwick, Coventry CV4
7AL, England}

\author{Thomas M. A. Fink}
\email{tmf20@cam.ac.uk}
\homepage{http://www.tcm.phy.cam.ac.uk/~tmf20/}
\affiliation{CNRS UMR 144, Institut Curie, 75005 Paris, France}
\affiliation{Theory of Condensed Matter, Cavendish Laboratory, Cambridge
CB3 0HE, England}

\author{Neill E. Bowler}
\email{Neill.Bowler@physics.org}
\homepage{http://uk.geocities.com/neill_bowler}
\altaffiliation[Present address: ]{Met Office,  Maclean Building,
Crowmarsh-Gifford, Oxfordshire OX10 8BB,  UK.}
\affiliation{Department of Physics, University of Warwick, Coventry CV4
7AL, England}

\begin{abstract}
We demonstrate that is it possible to simulate a system in thermal equilibrium
even when the energy cannot be evaluated exactly, provided the error
distribution is known. This leads to an effective optimisation strategy for
problems where the evaluation of each design can only be sampled statistically.

\end{abstract}


\maketitle

%%backup copy of addressdata
%\author{Robin C.\ Ball}
%\email{r.c.ball@warwick.ac.uk}
%\homepage{http://www.phys.warwick.ac.uk/theory/}
%\affiliation{Department of Physics, University of Warwick, Coventry CV4
%7AL, England}
%\author{Thomas M.\ A.\ Fink}
%\email{fink@lps.ens.fr}
%\homepage{http://www.tcm.phy.cam.ac.uk/~tmf20/}
%\affiliation{Theory of Condensed Matter, Cavendish Laboratory, Cambridge
%CB3 0HE, England}
%\affiliation{Morphog\'{e}n\`{e}se Cellulaire et Progression Tumorale,
%CNRS UMR 144, Institut Curie, 75005 Paris}
%\author{Neill E.\ Bowler}
%\email{Neill.Bowler@metoffice.com}
%\affiliation{Department of Physics, University of Warwick, Coventry CV4
%7AL, England}
%\presentaddress{Met Office,  Maclean Building,
%Crowmarsh-Gifford, Oxfordshire OX10 8BB,  UK.}
%%end address backup

We consider thermal equilibrium simulation of systems in which the energy of
any given state is either not known exactly, or else can much more cheaply be
estimated.\ A classic example which occurs in Physics is where each energy
calculation itself involves sampling over a distribution or numerical
integration, or the estimation of parameters of a numerical model. \ In this
letter we show how, by Stochastic Annealing, thermal equilibrium distributions
can nevertheless be sampled exactly, \ the essence of our method being that
the energy errors can be precisely absorbed as a contribution to thermal noise.

Our thermal sampling technique can be applied to optimisation problems where
the objective function is analogously difficult to evaluate, using Simulated
Annealing\cite{Kirkpatrick83} (meaning\ simulated cooling) or related methods
\cite{Harju97,Anderson01}. \ As an example we consider designing model protein
molecules to fold as fast as possible, \ where the only way to evaluate a
particular design is to run a sample of folding simulations. Confronted by
similar problems others have developed more empirical methods
\cite{robbins,gong,bulgak}, but none of these is underpinned by simulation of
true thermal equilibrium.

In a thermal ensemble the probability of the system occupying a state $\mu$
with energy $E(\mu)$ is
\begin{equation}
P(\mu)\propto e^{-\beta E(\mu)} \label{boltzmann system}%
\end{equation}
where $\beta=\frac{1}{T}$ is the inverse temperature. \ It is convenient to
sample this distribution by a Markov process, in which the system is allowed
to make a transition (move) from one state to another with rate constant
$K(\mu\rightarrow\nu)$ which depends only on the two states concerned. \ This
is generally more efficient than trying to choose the states directly,
\ provided we can assume that the move set is ergodic, \ meaning that all
states can be reached (eventually) from any given starting state. \ The more
strict condition of detailed balance,
\begin{equation}
P(\mu)K(\mu\rightarrow\nu)=P(\nu)K(\nu\rightarrow\mu),
\label{basic detailed balance}%
\end{equation}
imposes the correct equilibrium distribution provided
\begin{equation}
\frac{K(\mu\rightarrow\nu)}{K(\nu\rightarrow\mu)}=e^{-\beta\,\Delta E}
\label{detailed balance}%
\end{equation}
where $\Delta E$ is the energy difference $E(\nu)-E(\mu)$.

\begin{figure}[ptb]
\centering\resizebox{0.9\columnwidth}{!}{\includegraphics{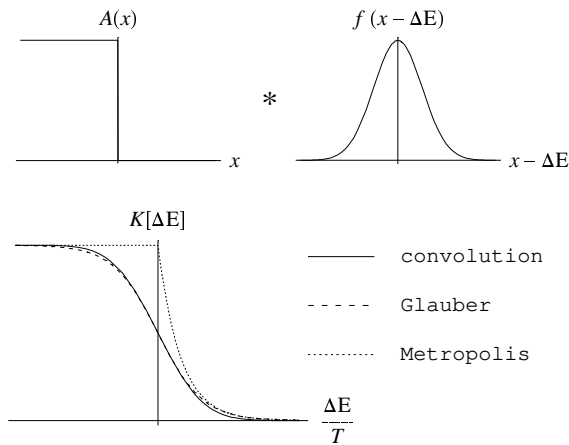}}\caption{A\ simple
approximate stochastic annealing is obtained by accepting moves on the basis
of the sign of the estimated energy change $x$. \ The acceptance probability
as a function of the true underlying energy change $\Delta E$ is then given by
a convolution with the error distribution. \ As shown for the Gaussian
distribution case, \ this gives an excellent approximation to the exact
Glauber acceptance rule.}%
\label{fig:convolution}%
\end{figure}

Whilst eq. (\ref{detailed balance}) ensures thermal equilibrium at inverse
temperature $\beta$, it does not fully determine the form of $K$. \ Typically
$K(\mu\rightarrow\nu)$ is the combination of an attempt frequency to move to
state $\nu$ given that the system is in state $\mu$, \ multiplied by an
acceptance probability. \ For simplicity of exposition we will take all the
attempt frequencies to be equal to unity, \ so that $K$ is just an acceptance
probability and so must obey $0\leq K\leq1$. The Metropolis
algorithm\cite{metropolis} is fully specified by requiring that $K(\Delta
E)=1$ for $\Delta E<C$, with $C$ as large as possible (maximising acceptance
rates) leading to
\begin{equation}
K_{\text{Metropolis}}(\Delta E)=\min\left(  1,e^{-\beta\Delta E}\right)
,\label{metropolis}%
\end{equation}
whereas the Glauber acceptance function \cite{glauber} arises by requiring
that $K(\Delta E)+K(-\Delta E)=1$, leading to
\begin{equation}
K_{\text{Glauber}}(\Delta E)=1/\left(  1+e^{\beta\Delta E}\right)
.\label{glauber}%
\end{equation}
These are compared graphically in Fig. \ref{fig:convolution}.

We now consider the case when the true energy change is not known exactly,
\ and we must accept moves with probabilty $A(x)$ where $x$ is only an
\emph{estimate} of the energy change. We will assume that these estimates have
statistically independent errors. \ If $f(x|\Delta E)$ is the probability
density of estimating that the energy change is $x$ when its true value is
$\Delta E$, \ then the net probability of accepting a move whose true energy
change is $\Delta E$ is given by
\begin{equation}
K(\Delta E)=\int_{-\infty}^{\infty}f(x\mid\Delta E)\,A(x)\,dx.
\label{probability of acceptance}%
\end{equation}
It is our aim to choose $A(x)$ such that $K(\Delta E)$ satisfies detailed
balance (\ref{detailed balance}).

\begin{figure}[ptb]
\centering\resizebox{0.9\columnwidth}{!}{\includegraphics{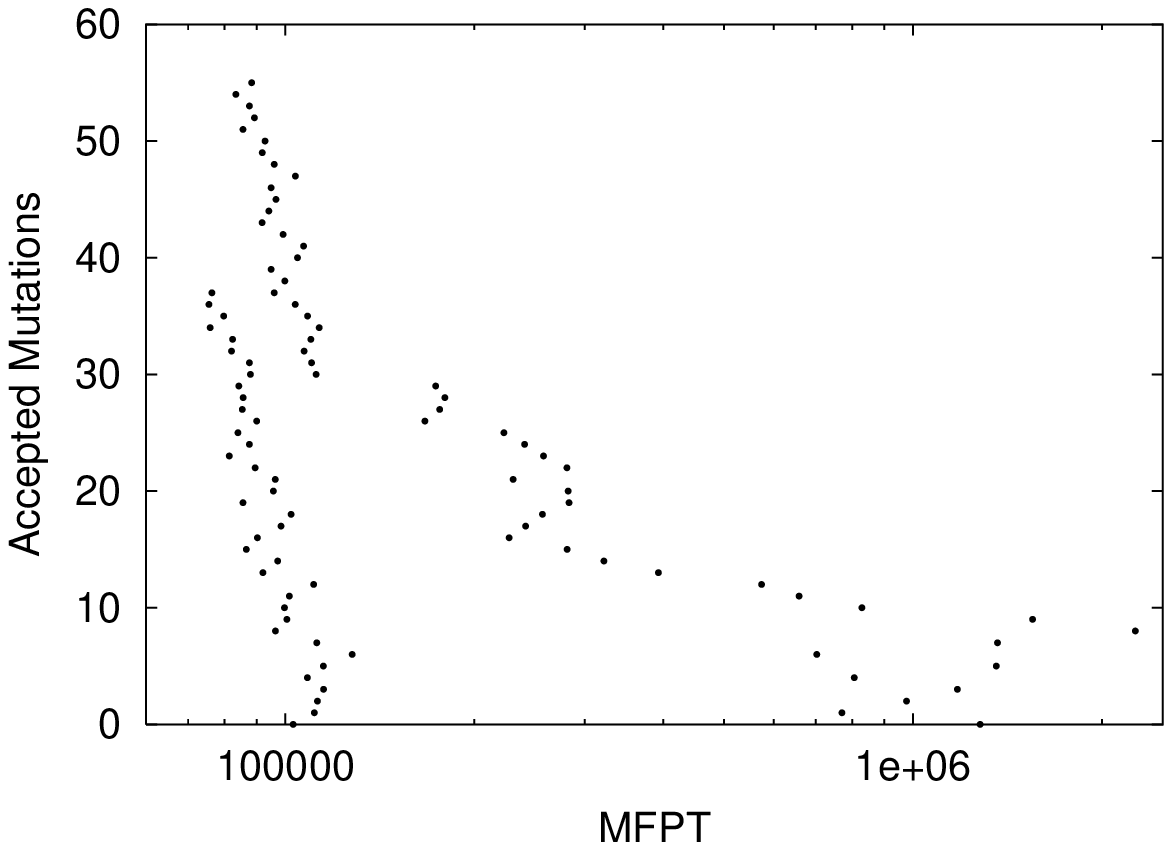}}
\centering\resizebox{0.9\columnwidth}{!}{\includegraphics{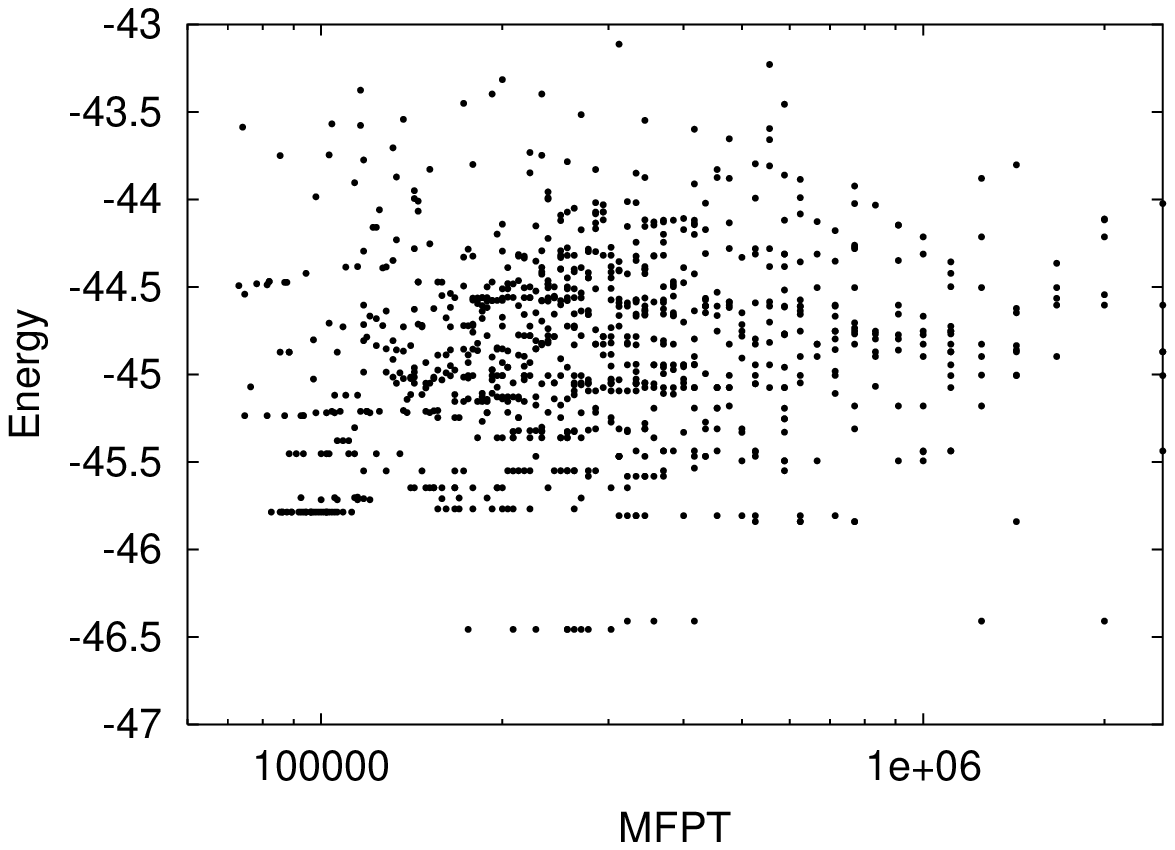}}\caption{Stochastic
annealing applied to protein folding, in a simple cubic lattice model. \ The
chains were 27 units long and the space of amino acid sequences was explored
for the ability of the molecule to fold spontaneously to a fixed
$3\times3\times3$ cubic target conformation. \ (a) The upper panel shows the
result of two stochastic annealing simulations in which successive mutations
of a starting sequence were accepted if a limited sampling of the Mean First
Passage Time (to the target) was improved. \ Increasing the depth of sampling
as the annealing proceeded ($y$-axis) provided the analogue of lowering the
temperature. \ (b) \ For comparison the lower panel shows the result of an
extensive search over sequences guided by thermal stability (energy in the
target conformation). \ Clearly individiual runs of Stochastic Annealing find
folding speeds approaching the fastest available, \ and much better than would
be achieved by seeking minimum energy (the Shaknovitich scheme \cite{Shak95}%
).}%
\label{fig:folding}%
\end{figure}

We can gain insight by considering the crude choice $A(x)=1$ for $x<0$ and
$A(x)=0$ otherwise. \ This simple strategy can give a good aproximation to eq.
(\ref{detailed balance}) of great value in optimisation. \ The resulting
acceptance function $K$ when $f$ is given by a Gaussian distribution is simply
related to the standard Error Function and is graphed in
Fig.\ \ref{fig:convolution}. The resemblance between this and the Glauber
acceptance function (whose symmetry it shares) is striking, showing how the
random energy errors make the selection look thermal - although in this case
the match is not exact, \ so detailed balance is not strictly achieved. \ The
standard deviation $\sigma$ of the Gaussian distribution controls an
approximate effective temperature using this rule, \ as inverting eq.
(\ref{detailed balance}) gives
\begin{equation}
T\simeq\sqrt{\frac{\pi}{8}}\sigma\left(  1-0.018\left(  \Delta E/T\right)
^{2}+\text{order}\left(  \Delta E/T\right)  ^{4}\right)
.\label{effective temperature}%
\end{equation}
For many optimisation purposes the departure from detailed balance, due to the
energy dependent terms at large $\left\vert \Delta E/T\right\vert $, is not a
problem. \ Fig. \ref{fig:folding} shows how we successfully used this approach
to optimise model protein folding: \ for each design change considered\ we
obtained estimates of the change in mean folding time from a limited sample of
folding simulations, \ and\ as the annealing proceeded we gradually cooled the
system\ by increasing the sample sizes leading to reduced error size and
reduced temperatures through Eq \ref{effective temperature}. \ In another
paper\cite{bowler} we show that this approach can be exploited to unravel a
benchmark problem in stochastic optimisation, the probabilistic travelling
salesman problem \cite{Jaillet88, Jaillet93}. 

We now attempt directly to choose $A(x)$ properly, \ such that $K$ exactly
satisfies detailed balance, eq.\ \ref{detailed balance}, \ assuming that the
error distribution $f(x|\Delta E)$ is fully known. We first note that this
leads to some bounds on the behaviour of the right tail of $A(x)$ and a left
tail of $f(x|\Delta E)$.

Given that $K(\Delta E)\leq1$ even for negative arguments, \ the detailed
balance condition requires $\int_{-\infty}^{\infty}f(x\mid\Delta
E)\,A(x)\,dx\leq e^{-\beta\Delta E}$ and for large positive $\Delta E$ \ this
severely restricts all contributions to the LHS. \ First consider
$x\simeq\Delta E$ \ for which $f(x\mid\Delta E)$ is not expected to be small.
\ Then the exponential decay on this contribution must come from $A(x) $
falling off at least as fast as $e^{-\beta x}$ as $x\rightarrow\infty$,
\ which becomes important in our later analysis. \ Second consider
$x\lesssim0$, \ for which $A(x)$ cannot become small or we would be heavily
rejecting even moves which appear to be downwards in energy. \ Then the
exponential decay on this contribution must come from $f(x|\Delta E)$,
\ $x\lesssim0$, falling off at least as fast as $e^{-\beta\Delta E}$ for
$\Delta E\rightarrow\infty$. \ This latter sets a fundamental restriction on
our stochastic annealing: the probability for large energy increases to be
estimated as negative must fall off faster than a thermal Boltzmann factor.

Now we begin direct analysis of the detailed balance condition
(\ref{detailed balance}), \ inroducing the substitutions $f(x|\Delta
E)=e^{\frac{\beta}{2}(x-\Delta E)}g(x,\Delta E)$ and $A(x)=e^{-\frac{\beta
x}{2}}a(x)$ so that eqs. (\ref{detailed balance}) and
(\ref{probability of acceptance}) combined simplify down to
\begin{equation}
\int_{-\infty}^{\infty}a(x)\,\left(  g(x,\Delta E)-g(x,-\Delta E)\right)
\,dx=0. \label{a's and g's}%
\end{equation}
Without significant loss of generality we can try
\begin{equation}
a(x)=\int_{-\infty}^{\infty}h(\Delta E^{\prime})\,j(x,\Delta E^{\prime
})\,d\Delta E^{\prime} \label{spotted solution}%
\end{equation}
where $h$ and $j$ are functions to be chosen. Substituting this into equation
\ref{a's and g's} we find
\begin{equation}
\int_{-\infty}^{\infty}h(\Delta E^{\prime})\left[  k(\Delta E,\Delta
E^{\prime})-k(-\Delta E,\Delta E^{\prime})\right]  \,d\Delta E^{\prime}=0
\label{must satisfy}%
\end{equation}
where
\begin{equation}
k(\Delta E,\Delta E^{\prime})=\int_{-\infty}^{\infty}g(x,\Delta E)j(x,\Delta
E^{\prime})\,dx. \label{This is k}%
\end{equation}
Then eq.\ \ref{must satisfy} is satisfied when $h(\Delta E^{\prime})=h(-\Delta
E^{\prime})$ and $k(\Delta E,\Delta E^{\prime})=k(-\Delta E,-\Delta E^{\prime
})$ are even functions in the given sense. We have not succeeded in taking
this most general case significantly further, the hard part being to implement
$0\leq A(x)\leq1$ for a probability.

For further progress we now specialise to the case of invariant error
distributions, \ where $f(x|\Delta E)=f(x-\Delta E)$, meaning that the
distribution of error is independent of $\Delta E$ itself. \ For this case, we
can choose $j(x,\Delta E^{\prime})=g(x-\Delta E^{\prime})$ and then from
eq.\ \ref{This is k} we find $k(\Delta E,\Delta E^{\prime})$ is a suitably
even function. \ Then we have a solution to eq \ref{a's and g's} given by
\begin{equation}
a(x)=\int_{-\infty}^{\infty}h(\Delta E^{\prime})\,g(x-\Delta E^{\prime
})\,d\Delta E^{\prime} \label{a=hg}%
\end{equation}
provided $h(x)$, and correspondingly $\widetilde{h}(p)$ below, is even.

We now aim to choose $h$ to maximise the move acceptance rates, \ which are
governed by $A(x)$. \ We follow the Metropolis methodology in choosing $A(x)$
($=e^{-\frac{\beta x}{2}}a(x)$) to be identically $1$ below some threshold,
\ $x<C$. Then using the Fourier-Laplace transform defined by $\widetilde
{f}(p)=\int_{-\infty}^{\infty}e^{-px}\,f(x)\,dx,$ the transform of $a$
becomes
\begin{equation}
\widetilde{a}(p)=\int_{-\infty}^{C}e^{\frac{\beta x}{2}}e^{-px}\,dx+\int
_{C}^{\infty}b(x)e^{-px}\,dx \label{splitting a}%
\end{equation}
where $b(x)=A(x)e^{\frac{\beta x}{2}}$ for $x\geq C$, $b(x)=0$ for $x<C$. From
eq.(\ref{a=hg}) the transform of the new function $b(x)$ is given by
\begin{equation}
\widetilde{b}(p)=\widetilde{h}(p)\widetilde{g}(p)-\frac{e^{\left(  \frac
{\beta}{2}-p\right)  C}}{\frac{\beta}{2}-p}{.} \label{alpha 1}%
\end{equation}

The exponential bound we established on the right tail of $A(x)$ implies that
$\widetilde{b}(p)$ is bounded for all $\operatorname{Re}p>-\frac{\beta}{2}$,
\ whereas the integrability of $f(x)$ together with the exponential bound on
its left tail only imply that $\widetilde{g}(p)$ is bounded for $-\frac{\beta
}{2}\leq\operatorname{Re}p<\frac{\beta}{2}$. \ Thus $\widetilde{h}(p)$ must be
chosen to cancel both any divergences of $\widetilde{g}(p)$ in
$\operatorname{Re}p\geq\frac{\beta}{2}$ and the apparent pole at $p=\beta/2.$
We first turn to the Wiener-Hopf method \cite{morse} to define
\begin{equation}
\widetilde{g}(p)=\widetilde{g_{L}}(p)\widetilde{g_{R}}(p)
\end{equation}
where $\widetilde{g_{L}}(p)$ is bounded and non-zero for $\operatorname{Re}%
p<\frac{\beta}{2}$, and $\widetilde{g_{R}}(p)$ is bounded and non-zero for
$\operatorname{Re}p>\frac{\beta}{2}$ ; \ it also follows from the bounded
window for $\widetilde{g}(p)$ that $\widetilde{g_{R}}(p)$ is bounded for the
wider range $\operatorname{Re}p>-\frac{\beta}{2}$. \ \ Then by choosing
$\widetilde{h}(p)=\frac{B}{\left(  \left(  \beta/2\right)  ^{2}-p^{2}\right)
}\frac{1}{\widetilde{g_{L}}(p)\widetilde{g_{L}}(-p)}$ , \ we can ensure that
$\widetilde{h}(p)\widetilde{g}(p)=\frac{B}{\left(  \left(  \beta/2\right)
^{2}-p^{2}\right)  }\frac{\widetilde{g_{R}}(p)}{\widetilde{g_{L}}(-p)}$ is
duly bounded for $\operatorname{Re}p>-\frac{\beta}{2}$ except for the
(desired) pole at $p=\frac{\beta}{2}$. \ Choosing the constant $B=\beta
\frac{\widetilde{g_{L}}\left(  -\frac{\beta}{2}\right)  }{\widetilde{g_{R}%
}\left(  \frac{\beta}{2}\right)  }$ makes the residue of this pole cancel when
we reassemble the expression (\ref{alpha 1}) for $\widetilde{b}(p)$ to give
\begin{equation}
\widetilde{b}(p)=-\frac{e^{\left(  \frac{\beta}{2}-p\right)  C}}{\frac{\beta
}{2}-p}+\frac{\beta}{\left(  \left(  \beta/2\right)  ^{2}-p^{2}\right)  }%
\frac{\widetilde{g_{R}}(p)\widetilde{g_{L}}\left(  -\frac{\beta}{2}\right)
}{\widetilde{g_{R}}\left(  \frac{\beta}{2}\right)  \widetilde{g_{L}}(-p)}.
\end{equation}

This should be the optimal solution. \ We cannot incorporate any new factors
into $\widetilde{h}(p)$ because, being even, they would have to be bounded
both for $Re(p)>-\frac{\beta}{2}$ and for $Re(p)>\frac{\beta}{2}$, \ and we
would run up against the limitations of Liouville's Theorem. \ The parameter
$C$ might appear to be a remaining degree of freedom, \ but it drops out of
the acceptance function itself which is given directly in terms of
\begin{equation}
\widetilde{a}(p)=\widetilde{h}(p)\widetilde{g}(p)=\frac{\beta}{\left(  \left(
\beta/2\right)  ^{2}-p^{2}\right)  }\frac{\widetilde{g_{R}}(p)\widetilde
{g_{L}}\left(  -\frac{\beta}{2}\right)  }{\widetilde{g_{R}}\left(  \frac
{\beta}{2}\right)  \widetilde{g_{L}}(-p)}. \label{a=hg2}%
\end{equation}
The parameter $C$ drops out because it simply reflects the partition we
introduced in equation (\ref{splitting a}). The feature which we have not
strictly guaranteed is that $A(x)=1$ for $x<C$ rather than some different cutoff.

As an example of our approach above, consider the simple case where
\begin{equation}
f(x-\Delta E)=\frac{\gamma}{2}e^{-\gamma\left|  x-\Delta E\right|  }
\label{exponential distribution}%
\end{equation}
which leads to $\widetilde{g}(p;\gamma)=\frac{\gamma^{2}}{\left(  \gamma
-\frac{\beta}{2}-p\right)  \left(  \gamma+\frac{\beta}{2}+p\right)  }.$ The
choice of $\widetilde{g_{L}}$ and $\widetilde{g_{R}}$ is trivial by
inspection, giving $\widetilde{a}(p)=\frac{\beta}{\left(  \beta/2\right)
^{2}-p^{2}}\frac{\gamma+\beta}{\gamma}\frac{\gamma-\frac{\beta}{2}+p}%
{\gamma+\frac{\beta}{2}+p}.$ The inverse transformation can also be performed
by inspection to give
\begin{equation}
A(x)=\min\left(  1,\frac{\gamma^{2}-\beta^{2}}{\gamma^{2}}e^{-\beta x}%
+\frac{\beta^{2}}{\gamma^{2}}e^{-(\gamma+\beta)x}\right)  .
\label{exponential a}%
\end{equation}
We note that this acceptance function only remains positive for $x\gg0$ when
$\gamma\geq\beta$, \ which is precisely the limit of achievable stochastic
annealing discussed earlier, \ and the case $\gamma\rightarrow\infty$ also
duly recovers the straightforward Metropolis method. \ We have analysed other
simple cases such as a rectangular error distribution, \ the superposition of
two exponentials as above, \ and the multiple convolution of exponentials,
\ all leading to results of equivalent properties.

The analysis of a Gaussian error distribution turns out to be slightly
singular, \ in that its corresponding $\widetilde{g}(p)$ has no obvious
Wiener-Hopf factorisation. \ However we can approach it by considering the
case where $x$ is taken to be a sum of $N$ independent exponential-distributed
variables,\ the error distribution taking the solvable form of multiply
convolved exponentials and (by the Central Limit Theorem) approaching Gaussian
form as $N\rightarrow\infty.$ In this case $\widetilde{g}(p)=\widetilde
{g}(p;\gamma)^{N}$ with $\gamma=\sqrt{2N}/\sigma$, where $\sigma$ is the
standard deviation of $x$, \ leading to $\widetilde{a}(p)=\frac{\beta}{\left(
\beta/2\right)  ^{2}-p^{2}}\left(  \frac{\gamma+\beta}{\gamma}\frac
{\gamma-\frac{\beta}{2}+p}{\gamma+\frac{\beta}{2}+p}\right)  ^{N}%
\rightarrow\frac{\beta}{\left(  \beta/2\right)  ^{2}-p^{2}}e^{\beta
(p-\beta/2)\sigma^{2}/2}$ as $N\rightarrow\infty$ at fixed $\sigma$. \ The
corresponding acceptance function by inverse transformation is then given by
\begin{equation}
A(x)=\min\left(  1,e^{-\beta(x+\beta\sigma^{2}/2)}\right)
\label{gaussian acceptance rule}%
\end{equation}

The optimal acceptance rules found above all obey the bounds that $0\leq
A(x)\leq1$ required for a probability, but they only do so \textit{a postiori}
so it can be objected that our optimal method gives no guarrantee of this
outcome \textit{a priori}. \ We have discovered a general but sub-optimal
solution for the acceptance function which does assure the requirement. \ The
key to this solution is to note that, \ given the result \ref{a=hg2}, our
requirements on the factorisation of $\widetilde{g}(p)$ can be relaxed to
require only that $\widetilde{g_{R}}(p)$ is bounded for $\operatorname{Re}%
p>-\frac{\beta}{2}$ and that $\widetilde{g_{L}}(p)$ is non-zero for
$\operatorname{Re}p<\frac{\beta}{2}$. \ Then assuming\ that $f(x)=0$ for
$x<C$, an acceptable factorisation is given by $\widetilde{g_{R}}(p)=$
$\widetilde{g}(p)e^{pC}$, $\widetilde{g_{L}}(p)=$ $e^{-pC}$ leading directly
to $\widetilde{a}(p)=\widetilde{h}(p)\widetilde{g}(p)=\frac{\beta}{\left(
\left(  \beta/2\right)  ^{2}-p^{2}\right)  }\frac{\widetilde{g}(p)}%
{\widetilde{g}\left(  \frac{\beta}{2}\right)  }$, \ at which point we can let
$C\rightarrow-\infty$ so no significant new restriction has been imposed on
$f$. \ The resulting acceptance function by inverse transformation can be
expressed as the convolution
\begin{equation}
A(x)=\frac{1}{\widetilde{f}(\beta)}\int_{-\infty}^{\infty}K_{\text{Metropolis}%
}(x-y)\ e^{-\beta y}f(y)\ dy. \label{general rule}%
\end{equation}
It can be verified by direct substitution that this does obey the detailed
balance condition (\ref{detailed balance}), it is manifestly positive
definite, \ and it gives maximum acceptance $1$ as $x\rightarrow-\infty$.
\ Simple comparison shows the resulting acceptance probabilities are lower
than the optimal values which we have calculated these, \ and the key
difference is that an interval where $A=1$ is no longer imposed.

Given the wide impact of equilibrium statistical mechanics, \ we are confident
that our new method of exactly thermal stochastic annealing will find
significant direct application. \ The chief limitation is that the error
distribution must be known. \ In this context the Gaussian case is particulary
important as it can be approached simply through multiple sampling, \ although
its variance is also required. \ Estimating the variance from the same sample
does induce errors, \ but these can be quantified and become negligible as the
sample size becomes large\cite{NeillPhD}. It is a more open question whether
it is viable to estimate by measurement the full distribution of error and
apply this to eq. (\ref{general rule}).

Underpinned by the exact results our analysis provides a powerful new tool in
stochastic optimisation. \ Generally it will be sufficient and convenient to
use the approximately thermal method we presented, of simply accepting moves
which appear to be an improvement. \ Then all the benefits of simulated
annealing are obtained by deliberately using crude estimates for each
decision! \ \ In the protein folding problem we used limited samples of
folding time for these estimates, \ making the high temperature part of the
annealing computationally the cheapest. \ This interestingly complicates the
long noted problem of how to choose the optimum cooling schedule, \ because
cooling by reducing sample errors introduces computational costs growing as
$1/T^{2}$ per move.

NEB would like to thank BP and EPSRC for the support of a CASE award during
this research.

\bibliographystyle{prsty}
\bibliography{References}

\end{document}